\documentclass[prd,floats,nofootinbib,showpacs]{revtex4}
\usepackage{graphicx}
\usepackage{amsmath}
\usepackage{amssymb}

\def\be{\begin{equation}}
\def\ee{\end{equation}}
\def\bea{\begin{eqnarray}}
\def\eea{\end{eqnarray}}
\def \wc {w_{c,\text{eff}}}
\def \wv {w_{x,\text{eff}}}

\renewcommand{\a}{{A}}
\newcommand{\h}{\mathcal{H}}

\begin{document}

\title{Large-scale instability in interacting dark energy and dark matter fluids}

\author{ Jussi V\"{a}liviita, Elisabetta Majerotto, Roy Maartens}

\affiliation{Institute of Cosmology \& Gravitation, University of
  Portsmouth, Portsmouth PO1 2EG, UK}

\begin{abstract}

If dark energy interacts with dark matter, this gives a new
approach to the coincidence problem. But interacting dark energy
models can suffer from pathologies. We consider the case where the
dark energy is modelled as a fluid with constant equation of state
parameter $w$. Non-interacting constant-$w$ models are well
behaved in the background and in the perturbed universe. But the
combination of constant $w$ and a simple interaction with dark
matter leads to an instability in the dark sector perturbations at
early times: the curvature perturbation blows up on super-Hubble
scales. Our results underline how important it is to carefully
analyze the relativistic perturbations when considering models of
coupled dark energy. The instability that we find has been missed
in some previous work where the perturbations were not
consistently treated. The unstable mode dominates even if
adiabatic initial conditions are used. The instability also arises
regardless of how weak the coupling is. This non-adiabatic
instability is different from previously discovered adiabatic
instabilities on small scales in the strong-coupling regime.

\end{abstract}


\maketitle

\section{Introduction}

In the standard cosmological model, dark energy and dark matter
are the dominant sources in the evolution of the late universe.
They are currently only indirectly detected via their
gravitational effects, and this produces an important
degeneracy~\cite{Kunz:2007rk}. In particular, there could be a
coupling between dark energy and dark matter without violating
observational constraints. A coupling in the dark sector could
help to explain why the dark energy only comes to dominate after
galaxy formation. But some of these models may be ruled out by
instabilities that are not apparent in the background solution.

Various forms of coupling have been considered (see
e.g.~\cite{Copeland:2006wr,Amendola:2006qi,boehmer} and references
therein). A general coupling may be described in the background by
the energy balance equations of cold dark matter ($c$) and dark
energy ($x$),
\begin{eqnarray}
  \label{relation}
\rho_c'  &=& - 3\h\rho_c-aQ\,,\label{cc}\\
\rho_{x}' &=& - 3\h(1+w_{x})\rho_{x}+ aQ\,, \label{kg1}
\end{eqnarray}
where $w_x=P_x/\rho_x$, $\h=d\ln a/d\tau$ and $\tau$ is conformal
time, with $ds^2=a^2(-d\tau^2+d\vec{x}^{\,2}\,)$.  Here $Q$ is the
rate of energy density transfer, so that $Q>0~(<0)$ implies that
the direction of energy transfer is dark matter~$\rightarrow$ dark
energy (dark energy~$\rightarrow$ dark matter).

The density evolution in the dark sector deviates from the
standard case. We can use effective equation of state parameters
for the dark sector to describe the equivalent uncoupled model in
the background: writing $\rho_c' +3\h(1+ \wc)\rho_c=0 $ and
$\rho_x' +3\h(1+ \wv)\rho_x=0 $, we have
 \be \label{weff}
\wc={aQ \over 3\h\rho_c}\,, \quad\quad  \wv= w_x -{aQ \over 3\h
\rho_x}\,.
 \ee
When $Q>0$, we have $\wc>0$, so that dark matter redshifts faster
than $a^{-3}$, while $\wv< w_x$, so that dark energy has more
accelerating power. The opposite holds for $Q<0$. When $Q>0$, the
coupled dark energy can behave like an uncoupled ``phantom" model,
i.e., with $\wv<-1$, but without the usual problems associated
with phantom dark energy~\cite{Huey:2004qv}.

In order to avoid stringent ``fifth-force" constraints, we assume
that baryons ($b$) and photons ($\gamma$) are not coupled to dark
energy and are separately conserved, and we assume the same for
neutrinos ($\nu$). So the balance equation for fluid $A$ is
\begin{equation}\label{eq. cont bg}
\rho_{\a}' = -3\h(1+w_\a) \rho_{\a} + aQ_{\a}\,,
\end{equation}
with $ Q_b=Q_\gamma=Q_\nu=0$ and $Q_c=-Q=-Q_x \neq0$. The Friedman
equation is
 \be\label{fe}
\h^2={8\pi G\over
3}a^2(\rho_\gamma+\rho_\nu+\rho_b+\rho_c+\rho_x)\,.
 \ee
Once a form for $Q$ is given, the background dynamics are fully
determined by the above equations, and typically the analysis
focuses on the possibility of accelerating attractor solutions
(for recent work with further references, see
e.g.~\cite{Copeland:2006wr,boehmer}). The models may also be
tested against geometric observational constraints (see
e.g.~\cite{Copeland:2006wr}).

In the perturbed universe, there are subtleties and complications
that do not arise for the background dynamics.
\begin{itemize}
\item
Firstly, one needs a covariant form for the dark sector
energy-momentum transfer that holds in an inhomogeneous universe,
and reduces to the background form in a Friedman-Robertson-Walker
(FRW) universe. For example, if one uses the ansatz $Q =Q_0 a^n$,
then the background dynamics can be determined and the parameters
$Q_0,n$ can be constrained by geometric observations. However,
there is no covariant form for such ad hoc ansatzes, and therefore
one is unable to compute the perturbations -- no consistent
cosmological model can be constructed on the basis of such
ansatzes.

\item
Secondly, one needs to ensure that dark energy perturbations are
stable, i.e., $c_{sx}^2>0$ where $c_{sx}$ is the dark energy sound
speed (the speed at which fluctuations propagate). For a scalar
field model of dark energy, $c_{sx}^2=1$ follows without
assumptions~\cite{Garriga:1999vw}. But for fluid models as used
here, we need to impose $ c_{sx}^2>0$ by hand, so that the dark
energy fluid is effectively non-adiabatic.

The sound speed problem applies equally to uncoupled dark energy,
but since the coupling itself can introduce non-adiabatic modes,
the issue is even more important in the coupled case.

\end{itemize}

Here we consider a dark energy fluid, with $w_x=\,$const, that is
coupled to dark matter via a covariant energy-momentum transfer
four-vector $Q^{\mu}$, which reduces in the background to $\pm
a^{-1}Q\delta^\mu_0$, where $Q$ is a simple function of energy
density. We show that the gauge-invariant curvature perturbation
has a super-Hubble instability in the early radiation era, no
matter how small the coupling is, and even if adiabatic initial
conditions are used. This rules out these models. It appears that
constant-$w_x$ fluid models of dark energy, even with the
imposition of $c_{sx}^2=1$, are unstable to couplings with the
dark matter. The non-adiabatic large-scale instability that we
find is different from the small-scale instabilities in the strong
coupling and adiabatic regime that have been previously
discussed~\cite{Koivisto:2005nr,Bean:2007nx}.

In order to avoid the large-scale instability, $w_x$ must increase
sufficiently in the early radiation and matter eras. In other
words, the simple constant-$w_x$ fluid model cannot be extended to
the early radiation era. This is in contrast to the case of
uncoupled dark energy, where constant-$w_x$ fluid models are well
behaved in the primordial universe. A dynamically evolving
quintessence field will likely avoid the instability we find,
since $w_x$ typically does not remain constant back to the early
radiation era.

The plan of the paper is as follows. In Sec.~II, we present the
density and velocity perturbation equations for a general model of
dark energy and an arbitrary form of coupling to dark matter. We
pay special attention to the dark energy sound speed and pressure
fluctuations. In Sec.~III we present a simple physically motivated
coupling, and we derive analytical solutions for the perturbations
in the early radiation era, in the case $w_x=\,$const. These
solutions reveal the non-adiabatic super-Hubble instability, which
is confirmed by numerical solutions using a modified version of
CAMB~\cite{camb}. In Sec.~IV, we extend the analysis to another
simple class of background couplings, and we show that the same
instability persists. We conclude in Sec.~V.

\section{Density and velocity perturbation equations}
\label{sec:conversion}

The general perturbation equations for coupled fluids are given
in~\cite{Kodama:1985bj} and various subsequent papers. We follow
broadly the notation of~\cite{Malik:2002jb} and specialize to the
case of dark sector coupling. We pay particular attention to the
covariant form of the coupling and the correct treatment of
momentum transfer (which vanishes in the background).

Scalar perturbations of the flat FRW metric are given in general
by
\begin{equation}
\label{eq. gen gauge} ds^2=a^2\Big\{ -(1+2\phi)
d\tau^2+2\partial_iB\,d\tau dx^i + \Big[(1-2\psi)\delta_{ij}+
2\partial_i\partial_j E\Big]dx^idx^j\Big\}.
\end{equation}
The background four-velocity is $\bar u^\mu=a^{-1}\delta^\mu_0$,
and the $\a$-fluid four-velocity is
 \be\label{ua}
u^\mu_\a = a^{-1}\Big(1-\phi, \partial^i v_\a \Big)\,, \quad\quad
u_\mu^\a = a\Big(-1-\phi, \partial_i[ v_\a +B] \Big),
 \ee
where $v_\a$ is the peculiar velocity potential. The volume
expansion rate, which generalizes the Newtonian relation
$\theta=\vec\nabla \cdot \vec v$, is~\cite{Ma:1995ey}
 \be
\theta_\a=-k^2 (v_\a+B)\,.
 \ee

\subsection*{Energy-momentum tensors}

We choose $u_\a^\mu$ as the energy-frame four-velocity, i.e.,
there is zero momentum flux relative to $u_\a^\mu$, so that
$T_{\a\,\nu}^\mu u_\a^\nu = -\rho_\a u_\a^\mu$. Then the
$\a$-fluid energy-momentum tensor is
 \be
T_{\a\,\nu}^\mu =(\rho_\a+ P_\a)u_\a^\mu u_\nu^\a+
P_\a\delta^\mu{}_\nu +\pi^\mu_{\a\,\nu}\,,
 \ee
where $\rho_\a= \bar\rho_\a+ \delta\rho_\a$ and, $P_\a=\bar P_\a+
\delta P_\a$. The anisotropic stress $\pi^\mu_{\a\,\nu}$ is given
by
 \be
\pi^0_{\a\,\nu}=0\,,
 \quad\quad
\pi^i_{\a\,j}=\left(\partial^i
\partial_j -{1\over3} \delta^i{}_j\nabla^2\right) \pi_{\a}\,.
 \ee
The total (conserved) energy-momentum tensor is  $T^{\mu}_{\
\;\nu}=\sum T^\mu_{\a\,\nu}$, so that
 \be \label{tott}
(\rho+P)u^\mu u_\nu+P\delta^\mu{}_\nu+\pi^\mu{}_\nu+q^\mu u_\nu
+q_\nu u^\mu= \sum_{\a}(\rho_\a+P_\a)u_\a^\mu u^\a_\nu + \sum_{\a}
P_\a +\sum_{\a} \pi^\mu_{\a\,\nu}\,.
 \ee
Here $q^\mu$ is the total momentum flux relative to the total
four-velocity $u^\mu$. In general this four-velocity has the form
 \be \label{totu}
u^\mu = a^{-1}\Big(1-\phi, \partial^i v \Big).
 \ee
The choice of $v$ depends on how the total four-velocity is
defined.

It follows from Eqs.~(\ref{tott}) and (\ref{totu}) that $\rho=\sum
\rho_\a\,,~ P=\sum P_\a\,,~\pi^\mu{}_\nu=\sum
\pi^\mu_{\a\,\nu}\,,$ and the total momentum flux is
$q^i=a^{-1}\sum (\rho_\a+P_\a)
\partial^i v_\a - a^{-1}(\rho+P)\partial^iv $. Thus the total energy
frame ($q^i=0 $) is defined by
 \be\label{vef}
(\rho+P) v=\sum (\rho_\a+P_\a) v_\a\,.
 \ee
This is the choice of $v$ that we will use from now on.

\subsection*{Energy-momentum balance}

The covariant form of energy-momentum transfer
is~\cite{Kodama:1985bj,Dunsby:1991xk}
\begin{equation}
\label{eqn:energyexchange} \nabla_\nu T^{\mu\nu}_{\a } = Q^\mu_{\a
}\,, \quad\quad
 \sum_A Q^\mu_{\a } = 0\,.
\end{equation}
A general energy-momentum transfer can be split relative to the
total four-velocity as~\cite{Kodama:1985bj,Malik:2002jb}
 \be
Q_\a^\mu =Q_\a u^\mu + F_\a^\mu\,,~~ Q_\a=\bar{Q}_\a + \delta
Q_\a\,, ~~ u_\mu F_\a^\mu=0\,,
 \ee
where $Q_\a$ is the energy density transfer rate and $F_\a^\mu$ is
the momentum density transfer rate, relative to $u^\mu$. Then it
follows that $F_\a^\mu=a^{-1}(0,\partial^if_\a)$, where $f_\a$ is
a momentum transfer potential, and
\begin{eqnarray}
Q^{\a }_0 & = & -a\Big[ Q_\a(1+\phi) + \delta Q_\a
\Big],\label{eqn:Qenergy}
\\
Q^{\a }_i & = & a\partial_i\Big[ f_\a+ Q_\a (v+B)\Big].
\label{eqn:Qmomentum}
\end{eqnarray}
The perturbed energy transfer includes a metric perturbation term
$Q_\a\phi$, in addition to the perturbation $\delta Q_\a$. The
perturbed momentum transfer is made up of two parts: the momentum
transfer potential $Q_\a(v+B)$ that arises from energy transport
along the total velocity, and the intrinsic momentum transfer
potential $f_\a$. In the background, the energy-momentum transfer
four-vectors have the form
 \be\label{qb}
Q^\mu_{c}= a^{-1}(Q_c,\vec 0\,) =a^{-1}(-Q,\vec 0\,)=-Q^\mu_x\,,
 \ee
so that there is no momentum transfer.

Total energy-momentum conservation implies
 \be
0=\sum Q_\a=\sum \delta Q_\a= \sum f_\a\,.
 \ee
For each $\a$-fluid, Eq.~(\ref{eqn:energyexchange}) gives the
perturbed energy and momentum balance equations (in Fourier
space),
 \bea
{\delta\rho}_{\a}'+3{\cal H}(\delta\rho_{\a}+\delta P_{\a}) -
3\left(\rho_{\a}+P_{\a}\right)\psi' -
{k^2}\left(\rho_{\a}+P_{\a}\right)\left(v_\a+E'\right) &=&
aQ_{\a}\phi+ a\delta Q_{\a}\,, \label{eq. pert coupled}\\
\left[\left(\rho_{\a}+P_{\a}\right)\left(v_\a +B \right)
\right]'+4{\cal H}\left(\rho_{\a}+P_{\a}\right)\left(v_\a +B
\right) +\left(\rho_{\a}+P_{\a}\right)\phi +\delta
P_\a-\frac{2}{3}\frac{k^2}{a^2}\pi_\a &=& a Q_\a(v+B)+af_\a\,.
\label{eqn:Malik17}
 \eea

\subsection*{Sound speed and pressure perturbations}

The sound speed $c_{s\a}$ of a fluid or scalar field, labelled by
$\a$, is the propagation speed of pressure fluctuations in the
$\a$ rest frame~\cite{Kodama:1985bj,Bean:2003fb,Gordon:2004ez}:
 \be\label{ss}
c_{s\a}^2=\left. {\delta P_\a \over \delta\rho_\a}\right|_{\rm rf}
\,.
 \ee
For a scalar field $\varphi$, the rest frame is defined by the
hypersurfaces $\varphi=\,$const, orthogonal to the rest-frame
four-velocity $u^\varphi_\mu \propto \nabla_\mu\varphi$. Thus the
kinetic energy density in the rest frame is
$-{1\over2}\nabla_\mu\varphi
\nabla^\mu\varphi=\varphi'^{\,2}/(2a^2)$, while $\delta\varphi=0$
in the rest frame, so that $\delta V=0$, where $V(\varphi)$ is the
potential. The density and pressure perturbations are consequently
equal in the rest frame: $\delta\rho_\varphi
=\delta({1\over2}a^{-2}\varphi'^{\,2}+V)= a^{-2}\varphi' \delta
\varphi' =\delta({1\over2}a^{-2}\varphi'^{\,2}-V)= \delta
p_\varphi $. The sound speed is therefore equal to the speed of
light, independent of the form of $V(\varphi)$:
 \be\label{sfss}
\delta\varphi\big|_{\rm rf}=0~~\Rightarrow~~ c^2_{s\varphi}=1\,.
 \ee

We can define the ``adiabatic sound speed" for any medium via
 \be
c_{a\a}^2={{P}'_\a \over {\rho}'_\a}=w_\a+{{w}'_\a \over
{\rho}'_\a/\rho_\a} \,.
 \ee
For a barotropic fluid, $c_{s}^2=c_{a}^2$, and if $w=\,$const,
then $c_{a}^2=w$. By contrast, for a scalar field,
$c_{s\varphi}^2\neq c_{a\varphi}^2\neq w_\varphi$.

The fluid model for dark energy with constant $w_x$ is at face
value a barotropic adiabatic model. But if we treat the dark
energy strictly as an adiabatic fluid, then the sound speed $
c_{sx}$ would be imaginary ($c_{sx}^2=c_{ax}^2=w_x<0$), leading to
instabilities in the dark energy. In order to fix this problem, it
is necessary to impose $c_{sx}^2>0$ by hand~\cite{Gordon:2004ez},
and it is natural to adopt the scalar field value
Eq.~(\ref{sfss}). Thus
 \be\label{xss}
c_{sx}^2=1\,,\quad\quad c_{ax}^2=w_x=\,\mbox{const}\,<0\,.
 \ee
This is what is done in the CAMB~\cite{camb} and
CMBFAST~\cite{cfast} codes.

In the perturbation equations~(\ref{eq. pert coupled}) and
(\ref{eqn:Malik17}), we need to relate $\delta P_\a$ to
$\delta\rho_\a$ via Eq.~(\ref{ss}). The $\a$ rest frame (the zero
momentum gauge or comoving orthogonal gauge) is the comoving
($v_\a|_{\rm rf}=0$) orthogonal ($B|_{\rm rf}=0$) frame, so that
 \be
T^i_{\a\, 0}\big|_{\rm rf}=0=T^0_{\a\, i}\big|_{\rm rf}\,.
 \ee
We make a gauge transformation, $x^\mu \to x^\mu + (\delta\tau_\a,
\partial^i\delta x_\a)$, from the rest frame gauge to a general gauge:
 \be
v_\a + B=(v_\a+B)\big|_{\rm rf}+\delta \tau_\a\,, \quad\quad
 \delta P_\a=
\delta P_\a\big|_{\rm rf}-P_\a'\delta\tau_\a\,, \quad\quad
 \delta \rho_\a=
\delta \rho_\a\big|_{\rm rf}-\rho_\a'\delta\tau_\a\,.
 \ee
Thus $\delta\tau_\a=v_\a+B$, and substituting into the pressure
and density fluctuations, we obtain
 \bea
\delta P_\a &=& c_{a\a }^2\delta\rho_\a+
\big(c_{s\a}^2-c_{a\a}^2\big)\left[ \delta\rho_\a + {\rho_\a'
}\left(v_\a+B \right) \right] = c_{a\a }^2\delta\rho_\a+ \delta
P_{{\rm nad}\,\a}\,, \label{delp2}
 \eea
where $ \delta P_{{\rm nad}\,\a}$ is the intrinsic non-adiabatic
pressure perturbation in the $\a$-fluid.

Our result applies to both coupled and uncoupled fluids, where the
difference enters via the term $\rho'_\a$. This recovers the
expression for the uncoupled case
in~\cite{Bean:2003fb,Gordon:2004ez}. For the coupled case, the
background coupling $Q_\a$ enters $\delta P_\a$ explicitly:
 \be\label{delp3}
\delta P_\a = c_{s\a }^2\delta\rho_\a +
\big(c_{s\a}^2-c_{a\a}^2\big)\Big[ 3\h(1+w_\a)\rho_{\a} -aQ_\a
\Big]{\theta_\a \over k^2}\,.
 \ee
This corrects the expression used in~\cite{Olivares:2006jr}, which
omits the $Q_\a$ term. As a consequence, there are errors in the
equations in~\cite{Olivares:2005tb,Olivares:2006jr} for
$\delta_x'$ and $\theta_x'$. We will discuss the implications of
this in Sec.~V.

\subsection*{General equations}

From the above equations we can derive evolution equations for the
dimensionless density perturbation
$\delta_\a=\delta\rho_\a/\rho_\a$ and for the velocity
perturbation $\theta_\a$ (which has dimension of $k$):
 \bea
&&\delta_\a'+3{\cal H}(c_{s\a}^2-w_\a)\delta_\a
+(1+w_\a)\theta_\a+ 3{\cal H}\big[3{\cal
H}(1+w_\a)(c_{s\a}^2-w_\a)+w_\a' \big] {\theta_\a \over k^2}
\nonumber\\
&&~~-3(1+w_\a)\psi'+ (1+w_\a)k^2\big(B-E'\big) ={aQ_\a \over
\rho_\a}\left[ \phi-\delta_\a+3{\cal H}(c_{s\a}^2-w_\a) {\theta_\a
\over k^2} \right]
+{a\over \rho_\a}\, \delta Q_\a\,,\label{dpa}\\
&& \theta_\a'+{\cal H}\big(1-3c_{s\a}^2\big)\theta_\a- {c_{s\a}^2
\over (1+w_\a)}\,k^2\delta_\a +{2\over
3a^2(1+w_\a)\rho_\a}\,k^4\pi_\a -k^2\phi
\nonumber \\
&&~~~= {aQ_\a \over (1+w_\a)\rho_\a}\big[ \theta-
(1+c_{s\a}^2)\theta_\a \big] -{a\over (1+w_\a)\rho_\a} \,
k^2f_\a\,. \label{vpa}
 \eea

The curvature perturbations on constant-$\rho_\a$ surfaces and the
total curvature perturbation (on constant-$\rho$ surfaces), are
given by the gauge-invariant quantities
 \be \label{z}
\zeta_\a=-\psi-{\cal H}{\delta\rho_\a \over \rho_\a'}\,,
\quad\quad \zeta = -\psi-{\cal H}{\delta\rho \over \rho'} =\sum
{\rho_\a' \over \rho'}\,\zeta_\a\,.
 \ee
The total energy conservation equation leads to
 \be \label{zdot}
\zeta'=-{\h \over (\rho+P)}\,\delta P_{{\rm nad}}\,.
 \ee
The gauge-invariant relative entropy perturbation for any two
fluids is
 \be \label{relent}
S_{AB} = 3\h \left({\delta\rho_B \over \rho_B'}- {\delta\rho_\a
\over \rho_\a'}\right)=3(\zeta_A-\zeta_B).
 \ee

\section{A covariant model of dark sector coupling}

In order to apply the equations of the previous section, we need
to choose a model of the dark sector coupling via a covariant
choice of the transfer four-vector $Q^\mu_{c }=-Q^\mu_x$. For
example, a coupling model motivated by scalar-tensor theory
has~\cite{Wetterich:1994bg}
 \be\label{stc}
Q^\mu_{c}=-Q^\mu_{x}=\beta(\varphi)T^\nu_{c\,\nu} \nabla^\mu
\varphi\,,
 \ee
where $\varphi$ is the scalar field dark energy and $\beta$ is a
coupling function. Using this form, the perturbed energy transfer
$\delta Q_{c}=-\delta Q_x$ and momentum transfer $f_c=-f_x$ can be
calculated unambiguously. Note that $\nabla^\mu \varphi$ is
parallel to the dark energy four-velocity $u_x^\mu$, i.e.
$Q^\mu_{c}=-Q^\mu_{x} \propto u_x^\mu$.

For more phenomenological models of coupling, especially in the
context of fluid dark energy, it is not always clear what the
covariant form of the transfer should be. For example, consider
the background transfer models~\cite{Zimdahl:2001ar}
 \be\label{A}
Q= {\h \over a}(\alpha_c\rho_c + \alpha_x \rho_x) \,,
 \ee
where $\alpha_{c}$ and $\alpha_{x}$ are dimensionless constants.
One problem with these models is ambiguity under perturbation:
What is the covariant form of $Q_c^\mu=-Q^\mu_x$ that reduces to
Eq.~(\ref{A}) in the background? This has not been made explicit
in previous work~\cite{Olivares:2005tb,Olivares:2006jr}, as
pointed out in~\cite{Koivisto:2005nr}.

A further problem with Eq.~(\ref{A}) is the explicit presence of
the universal expansion rate $\h$. This is designed for
mathematical simplicity rather than physical motivation: one does
not expect the dark sector coupling at each event to depend on the
global behaviour of the universe, but to depend only on purely
local quantities. A generous interpretation is that the $\h$
factor is an approximation to the temperature-dependence of the
interaction rate.

Here we propose a covariant model of coupling that avoids these
problems:
 \be\label{C}
Q^\mu_{c}=-Q^\mu_x= \Gamma \,T^{\,\nu}_{c\,\,\nu}\, u_c^\mu =
-\Gamma \rho_c\, u_c^\mu \,,
 \ee
where $\Gamma$ is a constant interaction rate, $\rho_c$ is the
dark matter density in the inhomogeneous universe, and $u_c^\mu$
is the dark matter four-velocity. The notable features of this
phenomenological coupling model are:
\\(1)~The interaction rate $\Gamma$ is `local', i.e. it is determined
by local interactions and not by the universal expansion rate.
\\(2)~In the rest frame of the dark matter, there is no momentum
transfer. (By contrast, for Eq.~(\ref{stc}) the momentum transfer
vanishes in the dark energy rest frame.)
\\(3)~The case $\Gamma>0$ corresponds in the dark matter frame
to the decay of dark matter into dark energy. This opens the
possibility of an alternative approach to the coincidence problem:
instead of trying to achieve a constant nonzero ratio
$\Omega_c/\Omega_x$ on the basis of primordially existing dark
energy, one could try to build models where the dark energy
accumulates via the decay of dark matter, and dominates in the
late universe because the decay rate $\Gamma$ is small.

\subsection*{Background dynamics }

In the background, the coupling (\ref{C}) reduces to
Eq.~(\ref{qb}), with
 \be\label{Cb}
Q=\Gamma \rho_c\,.
 \ee
When $\Gamma>0$, this coincides with a special case of a model in
which superheavy dark matter particles decay to a quintessence
scalar field~\cite{Ziaeepour:2003qs}. It also has the same form as
simple models to describe the decay of dark matter into
radiation~\cite{Cen:2000xv}, or a curvaton field into
radiation~\cite{Malik:2002jb}. The background dynamics with
Eq.~(\ref{Cb}) have been analysed in~\cite{boehmer} for the case
of scalar field dark energy. The growth factor and weak lensing
have been investigated for the case $ w_x(a)=w_0+w_a(1-a)$
in~\cite{Schaefer:2008ku}.

We impose the condition $w_x>-1$ so as to avoid a phantom fluid
model of dark energy. In order to have a close-to-standard matter
dominated era for structure formation, and in order to be
consistent with the observed angular diameter distance to last
scattering, it is necessary that $|\wc|$ is small, i.e.
 \be\label{qlim}
{|Q| \over \rho_c} \lesssim 0.1 H_0\,,
 \ee
where $H_0 = \h_0/a_0=\h_0$ is today's Hubble rate.

For this coupling model, Eq.~(\ref{weff}) implies that
 \be
\wc ={a\Gamma \over 3\h }\,, \quad\quad \wv - w_x=-{a\Gamma \over
3\h }\,{\rho_c \over \rho_x}\,,  \label{wcwx}
 \ee
where $|\Gamma|\ll H_0$ by Eq.~(\ref{qlim}). Since $w_x>-1$, it
follows that the total effective equation of state satisfies
$w_{\rm tot}>-1$, so that $H$ is a decreasing function and
therefore $a|\Gamma|/\h$ decreases as we look backward into the
past. Thus $\wc\ll 1$ for all times up to the present, i.e., the
dark matter effectively does not see the coupling for all times
from today to the past:
 \be\label{cdm}
\rho_c= \rho_{c0}a^{-3}\,.
 \ee
If $|\Gamma| / H_0 <3(1+w_x)\Omega_{x0} / \Omega_{c0}$, then the
dark sector coupling is negligible at late times, by
Eqs.~(\ref{wcwx}) and (\ref{cdm}), and we have
$\rho_x=\rho_{x0}a^{-3(1+w_x)}$. This is valid to the past until
the coupling term $|a\Gamma\rho_c|$ is equal to the redshift term
$|3\h(1+w_x)\rho_x|$ in Eq.~(\ref{kg1}). At earlier times, the
coupling term will dominate for a small enough $a$, regardless of
how small $|\Gamma|$ is.

In the radiation era,
 \be\label{Crad2}
{\cal H}=\tau^{-1}\,, \quad\quad a^2=H_0^2\Omega_{r0} \tau^2\,,
 \ee
and the energy balance equations~(\ref{cc}) and (\ref{kg1}) lead
to a simple solution for early times in the case $w_x<-2/3$:
 \be \label{Crad}
a\Gamma{\rho_c \over \rho_x} =(3w_x+2) \tau^{-1}\,, \quad\quad
w_x<-{2 \over 3} \,.
 \ee
From now on we assume that $w_x<-2/3$, which is consistent with
observations. Equations~(\ref{cdm}) and (\ref{Crad}) imply that
 \be
\rho_x'=-\h \rho_x~ \Rightarrow~ \wv=-{2 \over 3} ~~\mbox{and}~~
\rho_x\propto a^{-1}\propto \tau^{-1} \,. \label{radrx}
 \ee
Equation~(\ref{cdm}) shows that $\rho_c>0$ for all times.
Therefore by Eq.(\ref{Crad}), $\rho_x$ becomes
negative\footnote{%
Note that we can avoid $\rho_x<0$ for $\Gamma>0$ if $w_x>-2/3$ or
$w_x<-1$, but we exclude these cases.
} %
when $\Gamma>0$. This is the case that corresponds to the decay of
dark matter to dark energy. The rigidity of the assumption that
$w_x$ is constant leads to this problem with the decaying dark
matter case. When dark energy is modelled as a scalar
field~\cite{boehmer}, it remains positive for all times and there
is no such problem in the case $\Gamma>0$.

These analytical approximations are confirmed by numerical
integration, as illustrated in Fig.~\ref{fig:modAbgandpert}.

\subsection*{Dark sector perturbations }
\label{sec:perturbationeqns}

For the model described by the covariant energy-momentum transfer
four-vector $Q_c^\mu=-Q^\mu_x=-\Gamma \rho_c u_c^\mu$, we need to
determine $\delta Q_c=-\delta Q_x$ and $f_c=-f_x$ from the
conditions imposed by energy-momentum balance. Using
Eqs.~(\ref{ua}) and (\ref{C}), we find the components of
$Q^c_\mu=-Q^x_\mu$:
 \be
Q^c_\mu =-Q^x_\mu = a\Gamma\rho_c\Big[ 1+\phi+ \delta_c\,,~
\partial_i\left(v_c+B \right) \Big].\label{Qc}
 \ee
Comparing with Eqs.~(\ref{eqn:Qenergy}) and (\ref{eqn:Qmomentum}),
it follows that
 \be
\delta Q_c=-\Gamma\rho_c \delta_c= -\delta Q_x\,, \quad\quad f_c=
\Gamma\rho_c (v-v_c) = -f_x\,,
 \ee
where $v$ is the total energy frame velocity, defined by
Eq.~(\ref{vef}).

The density and velocity perturbation equations~(\ref{dpa}) and
(\ref{vpa}) for the dark sector, with $\pi_{c}=0=\pi_x$,
$w_{c}=0=w_x'$ and $c_{sx}^2=1$, can then be given, in
longitudinal (Newtonian) gauge ($B=E=0$):
 \bea
\delta'_x + 3\mathcal H(1-w_x)\delta_x + (1+w_x)\theta_x +
9\mathcal H^2(1-w_x^2)\frac{\theta_x}{k^2}-3(1+w_x)\psi'&=&
a\Gamma \frac{\rho_c}{\rho_x}\left[\delta_c -\delta_x + 3\mathcal
H (1-w_x)\frac{\theta_x}{k^2}+\phi\right]\!, \label{dpx}\\
\theta'_x -2 \mathcal H\theta_x -\frac{k^2}{(1+w_x)}\delta_x -
k^2\phi &=&
\frac{a\Gamma}{(1+w_x)}\frac{\rho_c}{\rho_x}\left(\theta_c-2\theta_x
\right), \label{vpx}
 \eea
and
 \bea
\delta'_c +\theta_c -3\psi' &=&- a\Gamma\phi\,, \label{dpc}\\
\theta'_c + \mathcal H \theta_c -k^2\phi &=& 0\,.\label{vpc}
 \eea
Note that the dark matter velocity perturbation
equation~(\ref{vpc}) is the same as in the uncoupled case. (In
particular, this means that in synchronous gauge, we can
consistently set $\theta_c=0$, as is done in the standard,
uncoupled case.) This is due to the fact that there is no momentum
transfer in the dark matter frame.

We are now in a position to see qualitatively why there is a
large-scale instability in the dark sector perturbations during
the early radiation era. The coupling term $Q_x$ in $\delta P_x$,
Eq.~(\ref{delp3}), leads to a driving term
\begin{equation}\label{qual}
-2\frac{a\Gamma}{(1+w_x)}\frac{\rho_c}{\rho_x}\theta_x =
\frac{-2(3w_x+2)}{1+w_x}\h\theta_x
\end{equation}
on the right hand side of Eq.~(\ref{vpx}). Here the multiplier of
$\h \theta_x$ is a positive number (since $w_x<-2/3$) -- and it
becomes very large if $w_x$ is close to $-1$. This causes rapid
growth of $\theta_x$. Qualitatively, this is the source of the
instability: in the presence of energy-momentum transfer in the
perturbed dark fluids, momentum balance requires a run-away growth
of the dark energy velocity. The precise form of the instability
is computed analytically below.

\subsection*{Radiation era}

Tight coupling between photons and baryons means that (a)~the only
nonzero momentum transfer is in the dark sector, and (b)~the only
nonzero anisotropic stress is that of the neutrinos (which have
decoupled). The perturbed Einstein equations reduce to
 \bea
3\tau^{-1} \psi' +k^2 \psi + 3\tau^{-2}\phi &=&- 4\pi G a^2
\delta \rho\,,\\
k^2(\psi' + \tau^{-1}\phi) &=& 4\pi G a^2 (\rho+P) \theta\,,
\label{myein-conb}\\
\psi''   +2 \tau^{-1}\psi'- \tau^{-2}\psi+ \tau^{-1}\phi' +
\frac{k^2}{3}(\psi-\phi) &=& 4\pi Ga^2\delta P\,,\\
\psi-\phi &=& 8 \pi G \pi_\nu\,.
 \eea
The perturbed balance equations in the dark sector are given by
Eqs.~(\ref{dpx})--(\ref{vpc}), with background coefficients
determined by Eq.~(\ref{Crad2}). For the photon-baryon sector,
 \bea
\delta'_{\gamma}&=&-\frac{4}{3}\theta_\gamma + 4 \psi'\,,
\quad\quad \delta'_b
=-\theta_b +3\psi'\,,\\
\theta'_\gamma &=& \frac{1}{4}k^2 \delta_\gamma + k^2 \phi\,,
\quad\quad \theta'_b = -{\cal H} \theta_b + c_{sb}^2 k^2 \delta_b
+k^2 \phi\,,
 \eea
and for neutrinos~\cite{Ma:1995ey},
\begin{eqnarray}
\delta'_\nu = -\frac{4}{3}\theta_\nu + 4\psi'\,, \quad\quad
\theta'_\nu = \frac{1}{4} k^2 \delta_\nu + k^2 \phi -
k^2\sigma_\nu\,, \quad\quad \sigma'_\nu =
\frac{4}{15}\theta_\nu\,,\label{nu}
\end{eqnarray}
where $\sigma_\nu := 2k^2\pi_\nu/ [3a^2(\rho_\nu+P_\nu)]$, and we
have neglected the neutrino octopole, i.e., we work to leading
order in $k\tau$.

\subsection*{Adiabatic initial conditions}

Now we look for a solution in the radiation era, in the
super-Hubble scale limit, $k\tau \ll 1$. We find that we can set
adiabatic initial conditions to lowest order in $k\tau$:
 \bea
&& \phi = A_\phi=\,\mbox{const}\,, \quad\quad \psi = \left(
1+{2\over 5}R_\nu \right)\phi\,, \quad\quad R_\nu:=
{\rho_\nu \over \rho_\nu+\rho_\gamma}\,,\label{adi1}\\
&& \delta_\gamma =\delta_\nu ={4\over 3}\delta_b ={4 \over 3}
\delta_c = -2\phi\,,\quad \quad
\theta_\gamma=\theta_\nu=\theta_b=\theta_c
={1\over2}(k\tau)\,k\phi\,, \quad\quad \sigma_\nu={1\over15}
(k\tau)^2
\phi\,, \label{adi2}\\
&& \delta_x={1\over 4}\delta_\gamma\,, \quad\quad \theta_x =
\theta_\gamma\,. \label{adi3}
 \eea
The expression for $\delta_x$ follows from
$\zeta_x-\zeta_\gamma=0$ [see Eq.~(\ref{relent})], using the
early-time attractor solution Eq.~(\ref{radrx}).

However, at higher order in $k\tau$, this solution is {\em not}
adiabatic. In the standard case with uncoupled dark energy, this
is not an issue -- since the deviation from adiabaticity is
decaying and suppressed~\cite{Doran:2003xq}. However, for the
coupled dark energy model considered here, the situation is
dramatically different -- because there is a strongly growing
non-adiabatic mode on super-Hubble scales. Even with adiabatic
conditions in the limit $k\tau\to 0$, the terms of higher order in
$k\tau$ contain the non-adiabatic mode, and this mode will
dominate since it is strongly growing (see below). Note that this
mode is regular, and it is stimulated by the dark sector coupling,
as explained via Eq.~(\ref{qual}).

The detailed analysis of all perturbative modes is given
elsewhere~\cite{prep}. There we also show numerically how the
total curvature perturbation starts off constant with the initial
conditions Eqs.~(\ref{adi1})--(\ref{adi3}), but begins to grow
dramatically after a short time (well before equality).

\subsection*{The dominant non-adiabatic mode}

In order to find the non-adiabatic mode, we assume a leading-order
power-law form for the perturbations:
 \be
\psi=A_\psi(k\tau)^{n_\psi}\,, \quad\quad
\phi=A_\phi(k\tau)^{n_\phi} \,, \quad\quad \delta_\a=B_\a
(k\tau)^{n_\a}\,, \quad\quad \theta_\a=C_\a(k\tau)^{s_\a}\,,
\quad\quad \sigma_\nu=D_\nu (k\tau)^{n_\sigma}\,,
 \ee
where $n_\psi$ and $n_\phi$ are not zero ($n_\psi=0=n_\phi$ is the
adiabatic case). The perturbed Einstein and balance
equations~(\ref{dpx})--(\ref{nu}) may then be solved, to leading
order in $k\tau$, in terms of $\psi$:
\begin{eqnarray}
\phi & = & J \psi\,,\\
\delta_\gamma = \delta_\nu & = & 4\psi\,,\\
\delta_b = \delta_c = \frac{3}{4}\delta_\gamma & = & 3\psi\,,
\label{erdel}\\
\theta_\gamma  =  \theta_\nu = \theta_b & = &\frac{(J+1)}
{(n_\psi+1)}
\,  (k\tau)\,k\psi\,,\\
\sigma_\nu & = &  \frac{4}{15(n_\psi+2)}\;
(k\tau)\,{\theta_\nu \over k}\,,\\
\theta_c & =& \frac{(n_\psi+1)}{(n_\psi+2)} \;
\frac{J}{(J+1)}\,\theta_\gamma\,,\\
\delta_x & = &   \frac{-2k^3(3w_x+2)}{\Gamma H_0^2\Omega_{c0}}
\frac{[n_\psi^2 + (J+1)n_\psi - (J+2)]}
{(n_\psi-1)}\, (k\tau)^{-3}\, \psi\,,\label{dxer}\\
\theta_x & = & -\frac{(n_\psi+2)}{3(1-w_x)} \, (k\tau)\,k
\delta_x\,. \label{eqn:earlythetax}
\end{eqnarray}
The gauge-invariant curvature perturbation, Eq.~(\ref{z}), is
given in terms of $\psi$ as
\begin{equation}
\zeta = -{1\over2}{(n_\psi +J+2)}\,\psi. \label{cper}
\end{equation}

The solution is thus fully determined up to an arbitrary
normalization of the amplitude parameter $A_\psi$. The stress
anisotropy parameter $J$ is given by
\begin{equation}
J:={A_\phi \over A_\psi} = 1 -
\frac{16R_\nu}{5(n_\psi+2)(n_\psi+1)+8R_\nu}\,.
\label{eqn:psiperphi}
\end{equation}
The power-law index $n_\psi$ is determined in terms of $w_x$ as
\begin{equation}
n_\psi =n_\pm= \frac{-(1+2w_x)\pm\sqrt{3w_x^2-2}}{1+w_x}\,.
\label{npsi}
\end{equation}
The fastest growing mode is the $n_+$-mode. Equation~(\ref{dxer})
shows that the modes are regular (i.e., well-behaved as $k\tau\to
0$) provided that Re$\,n_\pm \geq 3$. This leads to the
conditions,
 \be\label{reg}
n_+~\mbox{regular if}~-1<w_x \leq -{4\over 5}\,,~~~~
n_-~\mbox{regular if}~ -{9 \over 11}< w_x \leq -{4\over 5}
 \ee

The only explicit dependence on the coupling rate $\Gamma$ in the
early-time solutions is in the $\delta_x$ solution,
Eq.~(\ref{dxer}). The uncoupled limit $\Gamma=0$ leads to a
singularity in that equation. This reflects the fact that the
solutions are not valid in the limit $\Gamma = 0$. We cannot
recover the $\Gamma=0$ limit since we have used in an essential
way that $\Gamma \neq 0$, see Eqs.~(\ref{Crad}) and (\ref{radrx}).
The most important difference is that $n_\psi=0$ in the uncoupled
case, so that $\psi$ and $\phi$ are constant, and in addition
$\zeta=\,$const. By contrast, when $\Gamma \neq 0$, we see that
$n_+$ is typically very large:
 \be
w_x \sim -1~~ \Rightarrow ~~ n_+ \sim {2 \over 1+w_x} \gg 1\,.
 \ee
This large exponent signals the blow-up of $\psi$, and therefore
of all perturbations, including the gauge-invariant curvature
perturbation, Eq.~(\ref{cper}), on super-Hubble scales in the
early radiation era. The instability is stronger the closer $w_x$
is to $-1$.

The instability occurs no matter how weak the coupling is. A
smaller value of $|\Gamma|$ simply moves the blow-up to earlier
times. This is in contrast with the strong-coupling instabilities
discussed in~\cite{Koivisto:2005nr,Bean:2007nx}. Furthermore, the
instability is non-adiabatic, in accordance with Eq.~(\ref{zdot}),
since the curvature perturbation blows up. Again, this is in
contrast to the adiabatic instabilities
of~\cite{Koivisto:2005nr,Bean:2007nx}. The origin of this
large-scale non-adiabatic instability is not simply the fact that
the dark energy fluid is non-adiabatic, i.e., $c_{ax}^2 \neq
c_{sx}^2$. In the uncoupled case, the same non-adiabatic fluid
behaviour is also present, but there is no instability. The
coupling plays an essential role in driving the large-scale
non-adiabatic instability.

\begin{figure*}[!ht]
\centering
\includegraphics[width=0.48\textwidth]{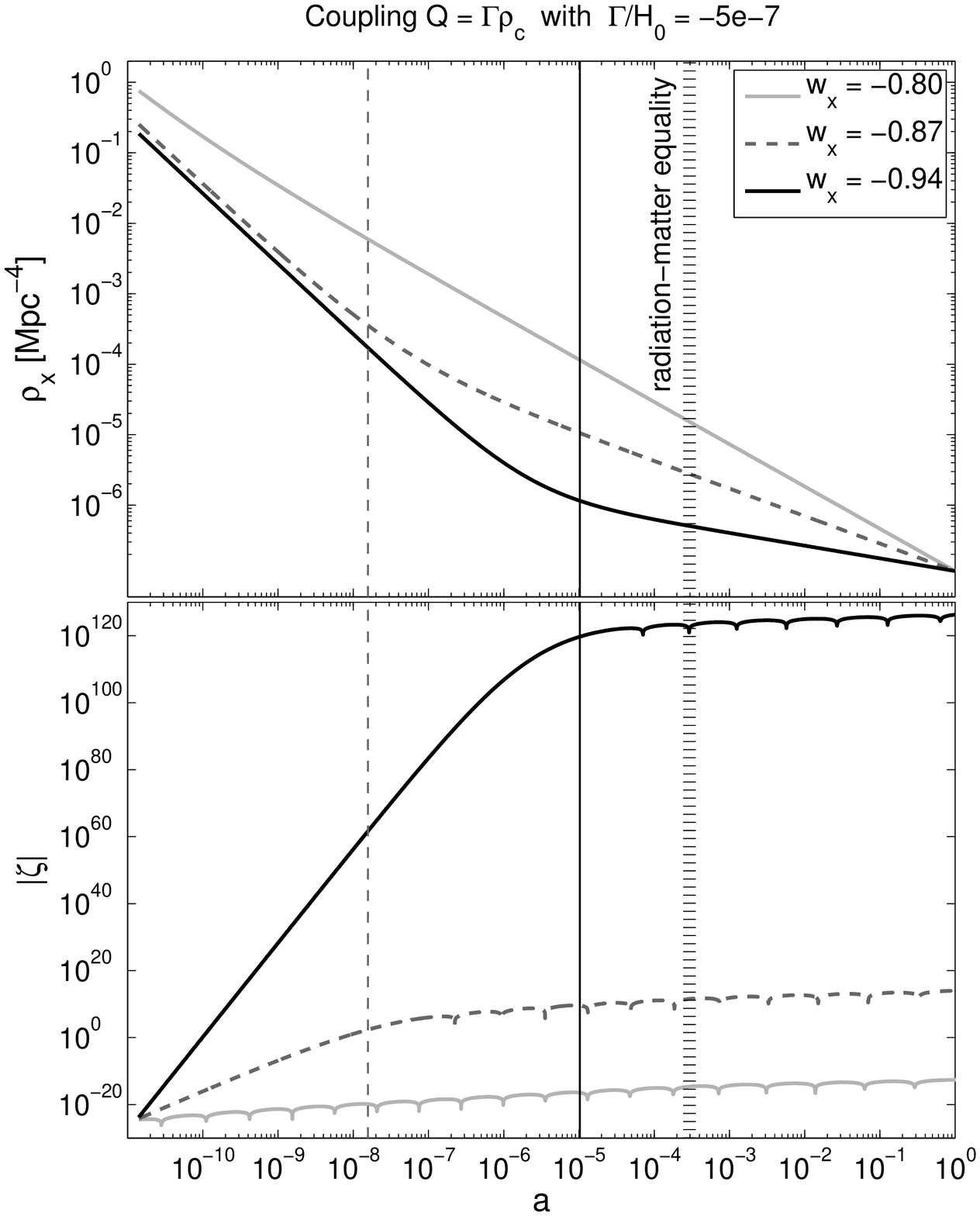}
\includegraphics[width=0.48\textwidth]{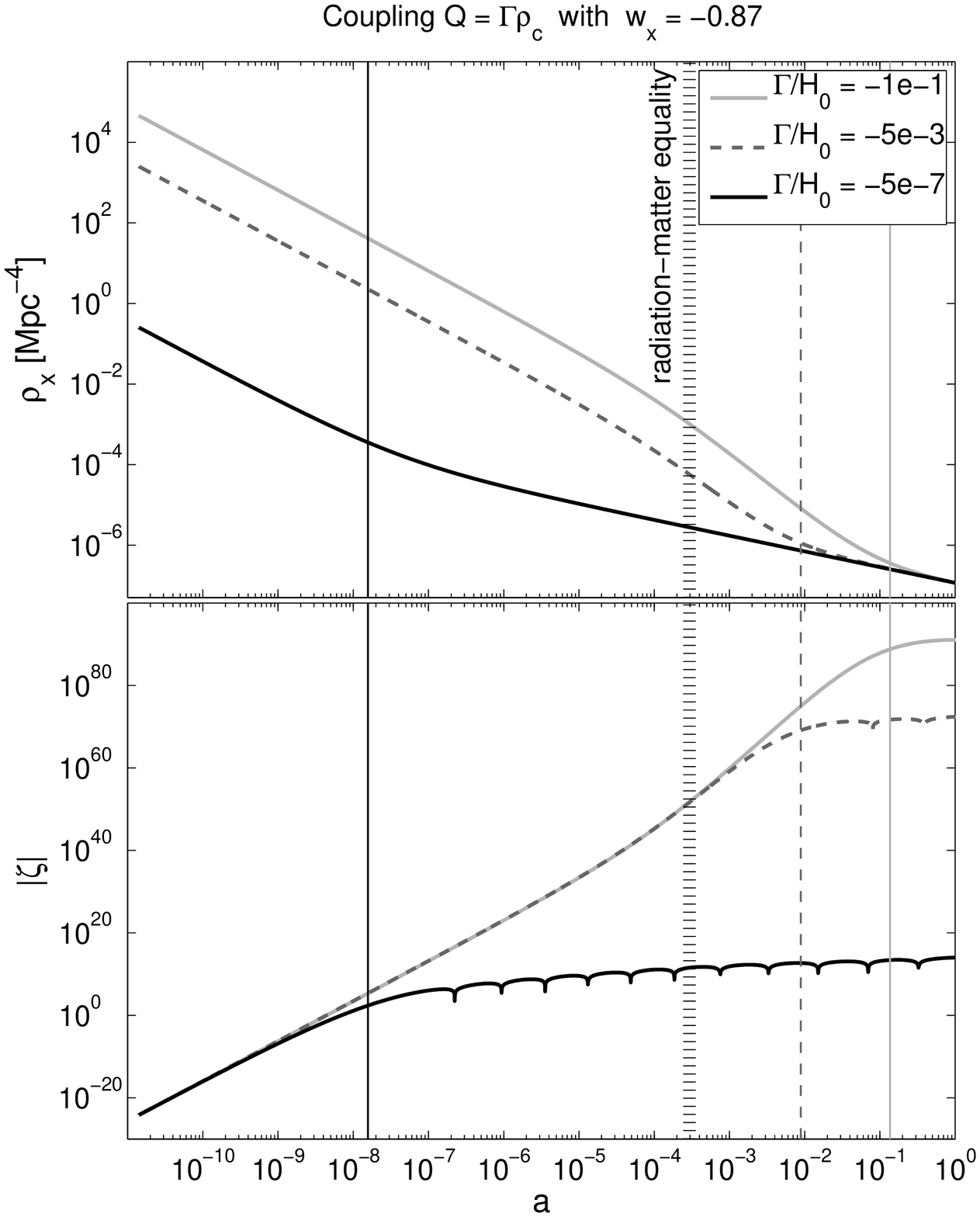}
\caption{The background evolution of dark energy density $\rho_x$
(top panels) and the evolution of the gauge-invariant curvature
perturbation $\zeta$ for a super-Hubble scale
$k=7\times10^{-5}\,$Mpc$^{-1}$ (bottom panels), as functions of
scale factor $a$, for the model with coupling given by
Eq.~(\ref{C}). In panels on the left, we vary $w_x$ with $\Gamma$
fixed, while the right-hand panels have fixed $w_x$ and varying
$\Gamma$. For each case, the vertical lines (in matching style)
indicate the moment when $|3\h(1+w_{x})\rho_{x}|$ and
$|a\Gamma\rho_c|$ are equal in Eq.~(\ref{kg1}). To the right of
these lines the background evolves as in the uncoupled case, i.e,
$\rho_x \propto a^{-3(1+w_x)}$. To the left, the coupling modifies
the background evolution to $\rho_x \propto a^{-1}$,
Eq.~(\ref{radrx}). (Note that in the left panels for $w_x=-0.80$
this happens very far in the past, not shown in the figure.) All
the curves show the full numerical solution obtained with our
modified version of CAMB, with the initial amplitude of $\psi$ set
to $10^{-25}$. The analytical solution for the blow-up of $\zeta$,
Eq.~(\ref{cper}), is practically indistinguishable from the
numerical solution at early times. \label{fig:modAbgandpert}}
\end{figure*}

The analytical demonstration of the large-scale instability at
early times is confirmed by numerical solutions, using a modified
version of CAMB. Examples are shown in
Fig.~\ref{fig:modAbgandpert}.  As long as the coupling modifies
the background evolution, the curvature perturbation is extremely
rapidly growing: $\zeta \propto a^{n_+}$ where $n_+$ is given by
Eq.~(\ref{npsi}) during radiation domination. When the background
starts to behave as uncoupled, the blow-up of $\zeta$ ends and it
begins to oscillate about zero with large (and mildly) increasing
amplitude.

Equation~(\ref{npsi}) shows that for large enough $w_x$, there are
oscillations super-imposed on the super-Hubble blow-up mode:
$-\sqrt{2 / 3}< w_x \leq -{4/ 5}$, where the upper limit comes
from Eq.~(\ref{reg}). See Fig.~\ref{fig:modAbgandpert} for an
example (left panel, with $w_x=-0.8$).

Figure~\ref{fig:modAbgandpert} also shows that if $|\Gamma|$ is
large enough to modify the background after radiation-matter
equality, then $|\zeta|$ follows a matter-dominated attractor
solution:
 \be
\zeta \propto a^{\hat{n}_+}~~~ \mbox{where}~~ \hat{n}_+ \sim
{3\over 2} n_+~~ \mbox{for}~~ w_x \sim -1\,.
 \ee
The background solution in the matter era that corresponds to
Eq.~(\ref{Crad}) is
 \be \label{mrad}
a\Gamma{\rho_c \over \rho_x} =3(2w_x+1) \tau^{-1}\,, \quad\quad
w_x<-{1 \over 2} \,.
 \ee

\begin{figure*}[!ht]
\centering
\includegraphics[width=0.48\textwidth]{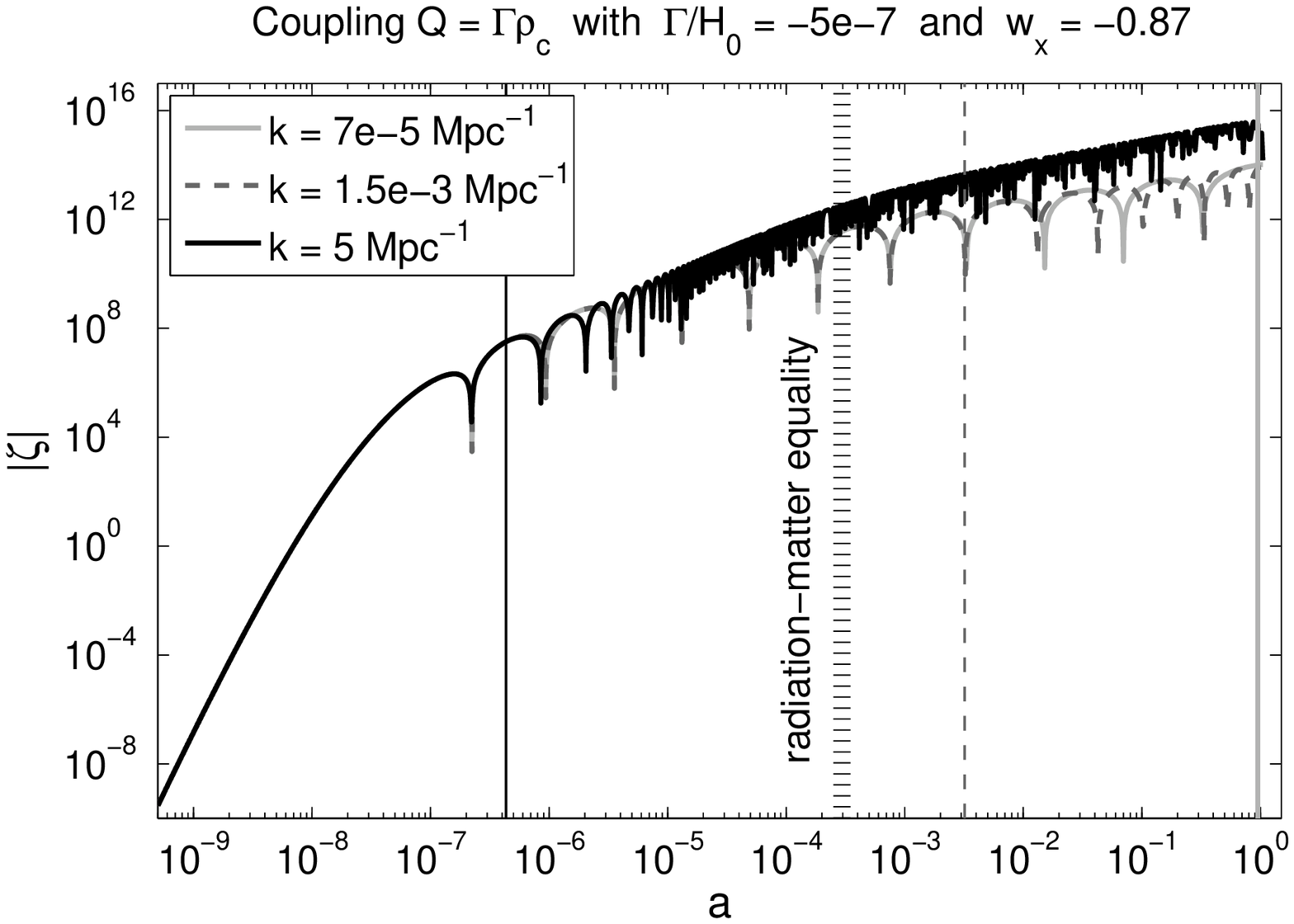}
\includegraphics[width=0.48\textwidth]{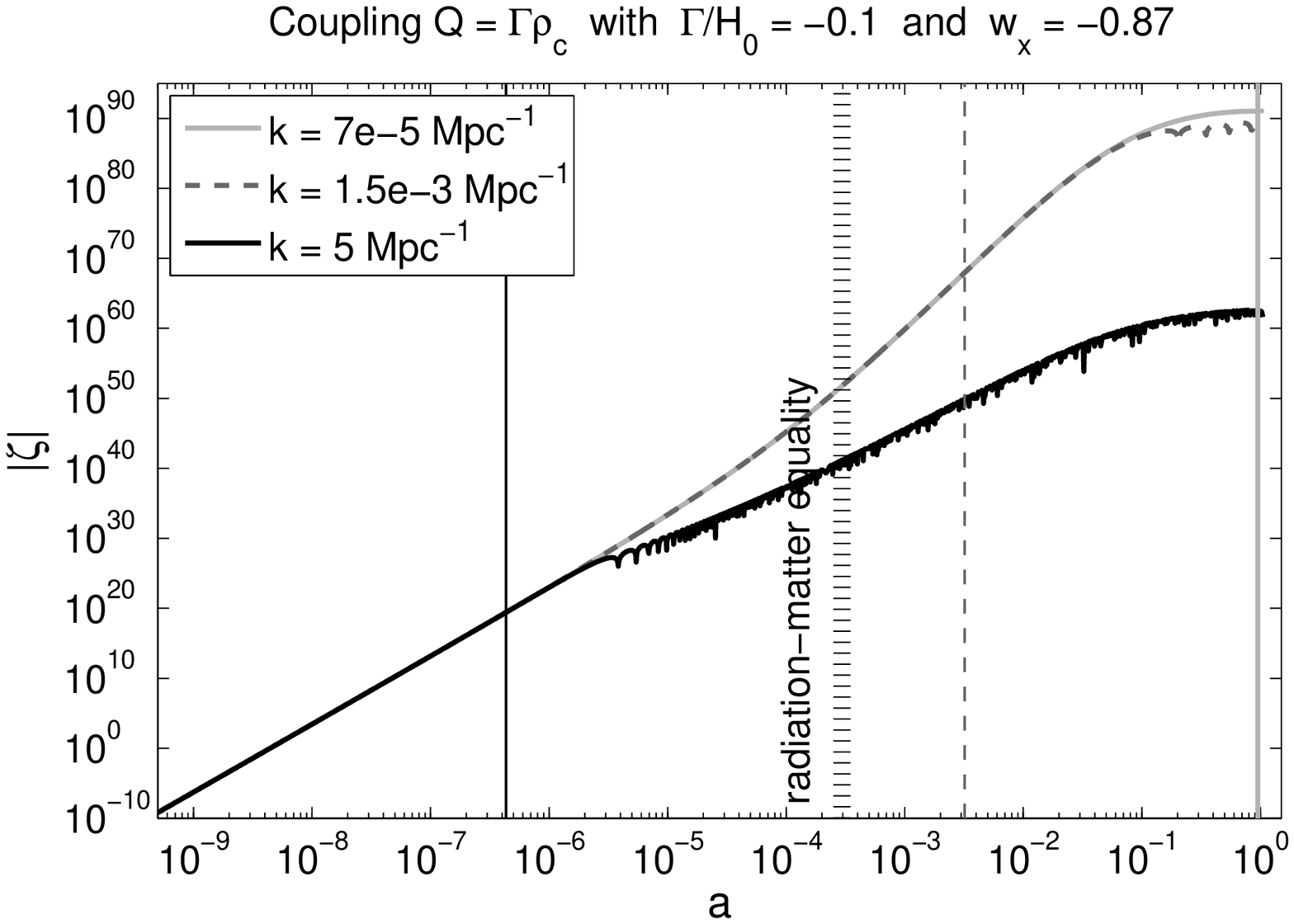}
\caption{The evolution of the gauge-invariant curvature
perturbation $\zeta$ for three different scales as a function of
scale factor $a$, for the model with coupling given by
Eq.~(\ref{C}). In the panel on the left, the coupling $|\Gamma|$
is very small while in the right-hand panel, $|\Gamma|$ is larger.
Vertical lines indicate the moment when each mode enters the
horizon ($k\tau \sim 1$). The largest scale ($k =
7\times10^{-5}\,$Mpc$^{-1}$) stays super-Hubble all the way up to
today. The intermediate scale ($k = 1.5\times10^{-3}\,$Mpc$^{-1}$)
enters the horizon during matter domination, and the smallest
scale ($k = 5\,$Mpc$^{-1}$) enters deep in the radiation
era.\label{fig:modAscaledep}}
\end{figure*}

The scale-dependence of the instability is illustrated in
Fig.~\ref{fig:modAscaledep}. The important point is that the
instability cannot be removed by simply re-scaling $A_\psi$: even
if we can match the large-scale CMB power with a small enough
$A_\psi$, the full CMB and matter power spectra will exhibit a
strong scale-dependence in violation of observations.

Note also that the analytical derivation of the instability does
not depend on the sign of $\Gamma$, i.e., there is a blow-up for
positive and negative $\Gamma$. However, as shown by
Eq.~(\ref{Crad2}), when $\Gamma>0$, there is always a time
$\tau_*$ when $\rho_x$ goes through zero and is negative for
$\tau<\tau_*$. The perturbation equations are singular at
$\tau_*$.

\section{Extension to other coupling models}

The instability in the model with coupling
$Q_c^\mu=-Q^\mu_x=-\Gamma \rho_c u_c^\mu$ is not peculiar to the
particular form of the coupling. The other models discussed in
Sec.~III also suffer from this instability. We consider two
background couplings, which are special cases of Eq.~(\ref{A}):
$aQ=\alpha \h \rho_c$ and $aQ=\beta \h (\rho_c+\rho_x)$.

\subsection*{Model with background coupling
$\, aQ=\alpha \h \rho_c$}

For this coupling the background balance equations may be solved
exactly:
 \bea
\rho_c &=& \rho_{c0}a^{-(3+\alpha)}\,,\label{A1sol1}\\
\rho_x &=& \rho_{x0}a^{-3(1+w_x)}+\left( {\alpha \over
\alpha-3w_x} \right)\rho_{c0}a^{-3}(a^{-3w_x}-a^{-\alpha})\,,
\label{A1sol2}
 \eea
where we assume that $\alpha >3w_x$ (otherwise the coupling
strength $|\alpha|$ would be too large). The dark matter density
is always positive. For the dark energy density to cross zero and
become negative, there must be a solution $a_{\rm n}<1$ to $a_{\rm
n}^{3w_x-\alpha}=1+ \Omega_{x0} (\alpha-3w_x)/( \Omega_{c0}
\alpha)$. The left hand side is always $>1$, since $3w_x-\alpha
<0$, whereas the right hand side is $>1$ only for $\alpha>0$. Thus
for $\alpha>0$, i.e. for the case of dark matter decaying into
dark energy, the dark energy density always becomes negative in
the past. In the early universe, $a\ll 1$, the exact solution
implies
 \be\label{Arad}
{\rho_x \over \rho_c} \to {\alpha \over 3w_x-\alpha}\,, \quad\quad
 \alpha >3w_x \,.
 \ee

To analyze perturbations in this model, we need a covariant form
of energy-momentum transfer four-vector that reduces to $Q=\alpha
\h\rho_c/a$ in the background. We propose to use the same form as
Eq.~(\ref{C}):
 \be\label{AQ}
aQ^\mu_c=-aQ^\mu_x =-\alpha {\h } \rho_c u_c^\mu\,.
 \ee
Here we are following the implicit assumption made in other work
that $\alpha H$ is an approximation to an interaction rate that
varies with time but not in space, so that there is no
perturbation of $H$ in $\delta Q^\mu_c$.

With this covariant form of energy-momentum transfer, the momentum
transfer is
 \be
af_c=\alpha {\h } \rho_c(v-v_c)=-af_x\,,
 \ee
and the dark sector density and velocity perturbations are given
by Eqs.~(\ref{dpx})--(\ref{vpc}), with $\Gamma$ replaced by
$\alpha \h/a$.

The early radiation solution to leading order in $k\tau$ is
qualitatively similar to the solution for the $\Gamma$ model in
the previous section, with differences arising because the
interaction rate $\alpha H$ varies with time, as opposed to the
constant rate $\Gamma$. The solutions for the dark energy
perturbations become:
 \bea
\theta_x &=& {2\sqrt{\Omega_{r0}}\,(3w_x-\alpha) \over 3\alpha
\Omega_{c0}(1+w_x)}\,{[n_+^2+(J+1)n_+ -(J+2)] \over (n_++1)}\,
{k^2 \over H_0}\,\psi\,,
\label{vpxB}\\
\delta_x &=&-{3(\alpha+3)(1-w_x)\over
(n_++2-\alpha)}\,(k\tau)^{-1}\, {\theta_x \over k}\,. \label{dpxB}
 \eea
The key indicator of instability, i.e. the power-law index for the
fastest growing mode, $n_+$, takes a more complicated form:
\begin{equation}\label{nB}
n_+  = \frac{3\alpha+(\alpha-6) w_x +\sqrt{(\alpha^2 +16\alpha+
40) w_x^2-2 (\alpha^2 +6\alpha+8) w_x+(\alpha^2- 4 \alpha-20)}}{2
(1+w_x)}\,.
\end{equation}
For example, if $w_x = -0.87$ and $\beta = -0.003$, then $n_+ =
38.95$. Unlike the $\Gamma$-model expression Eq.~(\ref{npsi}),
here $n_+$ depends explicitly on the interaction rate parameter
$\alpha$. Note that, as for the $\Gamma$-model, the limit
$\alpha=0$ is not admitted in Eqs.~(\ref{vpxB})--(\ref{nB}), since
the derivation uses $\alpha \neq 0$ in an essential way. For small
$|\alpha|$ and $w_x$ close to $-1$,
 \be
n_+ \sim \frac{6}{1+w_x}\gg 1\,.
 \ee
This is triple the corresponding index for the $\Gamma$ model. The
analytical form for the early-time instability is confirmed by
numerical integration.

\subsection*{Model with background coupling
$\, aQ=\beta \h (\rho_c+ \rho_x)$}

The background coupling $Q$ is proportional to the total dark
sector density $\rho_c+\rho_x$, which obeys an energy conservation
equation. The balance equations lead to an exact
solution~\cite{Zimdahl:2001ar} for $\rho_x/\rho_c$, with 3 cases
according to the sign of $-\beta+3w_x/4$. The non-negative cases
are not relevant since they violate $|\beta|\ll 1$. For the
remaining case,
 \be\label{A2sol}
{\rho_x \over \rho_c}={(B+2\beta-3w_x) \over 2\beta} \left(
{1-ba^{B+2\beta-3w_x} \over
1+ba^{B+2\beta-3w_x}}\right)+{3w_x-2\beta \over 2\beta}\,,
\quad\quad \beta>{3w_x \over 4}\,, \ee where \be
B:=\sqrt{3w_x(3w_x-4\beta)}-2\beta +3w_x\,,
 \ee
and $b:=[B- 2\beta\Omega_{x0}/\Omega_{c0}]/ [B-6w_x+
2\beta(2+\Omega_{x0}/\Omega_{c0})]$. It follows that
 \be\label{Brad}
{\rho_x \over \rho_c}\to {1\over 2\beta} \times \left\{
\begin{array}{lll}
B\, & & ~~a\to 0 \\
B-2\sqrt{3w_x(3w_x-4\beta)} & & ~~a\to \infty \end{array}\right.
 \ee
If $\beta>0$, then $\rho_x/\rho_c$ is negative in the early
universe, $a\to 0$, and also in the future, $a\to\infty$.
Therefore the $\beta>0$ case of this model is unphysical. For
$\beta<0$, if we fix $\Omega_{x0}/\Omega_{c0}$ at a value greater
than the late attractor in Eq.~(\ref{Brad}), then in the past
$\rho_x/\rho_c$ becomes negative. For a physical model, we thus
require $3w_x/4<\beta<0$ and $\Omega_{x0}/\Omega_{c0}$ less than
the late-time attractor.

Perturbation of this coupling model is more complicated because it
is determined by the total density, and therefore there is more
ambiguity in the appropriate choice of four-velocity in the
definition of $Q^\mu_c=-Q^\mu_x$. In previous
work~\cite{Olivares:2005tb,Olivares:2006jr}, this issue was not
explicitly discussed, and no form for $Q^\mu_c=-Q^\mu_x$ was given
(this was also pointed out in~\cite{Koivisto:2005nr}). It appears
that the dark sector perturbation equations
in~\cite{Olivares:2006jr} do not conform to momentum balance. They
also neglect the coupling term in the expression for $\delta P_x$,
i.e., the $Q_x$ term in Eq.~(\ref{delp3}), as we pointed out in
Sec.~II. It turns out that this second error is decisive for the
instability, whereas the error in $f_c=- f_x$ only leads to small
corrections.

Using the correct form, Eq.~(\ref{delp3}), for $\delta P_x$, the
evolution equation for dark energy density perturbations (which is
independent of $f_x$) becomes
 \bea
&& \delta'_x + 3 \mathcal{H} (1-w_x)\delta_x + (1+w_x) \theta _x +
9 \mathcal{H}^2 (1-w_x^2) \frac{\theta_x}{k^2} - 3 (1+w_x) \psi'
\nonumber\\ &&~~~~= \beta \mathcal{H} \left[\left(1+
\frac{\rho_c}{\rho_x}\right) \left\{\phi + 3 \mathcal{H} (1-w_x)
\frac{\theta_x}{k^2}\right\} +  \frac{\rho_c}{\rho_x} (\delta_c -
\delta_x )\right]
 \label{dpxC}
 \eea
In~\cite{Olivares:2005tb,Olivares:2006jr}, Eq.~(7), the right hand
side has $-\delta_c-\rho_x\delta_x/\rho_c$ instead of $\delta_c
-\delta_x$. We find an instability (see below), whereas there is
no instability in the results of~\cite{Olivares:2006jr}. Their
omission of the coupling term in $\delta P_x$ has inadvertently
removed the instability.

The dark matter density perturbations obey
 \be
\delta'_c + \theta_c - 3 \psi' = -\beta
\mathcal{H}\left[\left(1+\frac{\rho_x}{\rho_c}\right) \phi +
\frac{\rho_x}{\rho_c} (\delta_x - \delta_c)\right].
 \ee
The equations for $\delta_x'$ and $\delta_c'$ are independent of
the momentum transfer $f_c=-f_x$. In order to compute the velocity
perturbations, we need the momentum transfer. If we follow the
previous models and choose the energy-momentum transfer
four-vector to be aligned with the dark matter four-velocity (so
that there is no momentum transfer in the dark matter frame), then
\begin{equation}\label{QC}
aQ^{\mu}_c =-aQ^\mu_x= -\beta \h (\rho_c + \rho_x) u^{\mu}_{c}\,.
\end{equation}
It follows that
\begin{equation}
af_c = \beta \h (\rho_c + \rho_x)(v - v_c)=-af_x \,,
\end{equation}
and the velocity perturbation equations become
\begin{eqnarray}
\theta'_x - 2\mathcal{H}\theta_x -\frac{k^2}{(1+w_x)}\delta_x -k^2
\phi &=& \frac{\beta \mathcal{H}}{(1+w_x)}\left(1+ \frac{\rho_c}
{\rho_x}\right)\left[\theta_c-2 \theta_x\right]\,, \label{vpxC}
\\
\theta'_c + \mathcal{H}\theta_c -k^2 \phi &=& 0\,.\label{vpcC}
\end{eqnarray}
The $\theta_c'$ equation agrees
with~\cite{Olivares:2005tb,Olivares:2006jr}, but their $\theta_x'$
equation has $+\theta_x$ in place of our $\theta_c-2 \theta_x$ on
the right-hand side. Any other choice for the four-velocity along
$Q_c^\mu$ will lead to a nonzero right-hand side in
Eq.~(\ref{vpcC}). It appears that no consistent choice of
$Q^\mu_c$ can recover the equations
of~\cite{Olivares:2005tb,Olivares:2006jr}.

We find that the problem of not accounting correctly for momentum
transfer has a minor effect on the instability. The key driver for
the instability is the coupling term in $\delta P_x$, leading to
the correct forms Eq.~(\ref{dpxC}) and (\ref{vpxC}) for the
$\delta_x'$ and $\theta_x'$ equations.

\begin{figure*}
\centering
\includegraphics[width=0.48\textwidth]{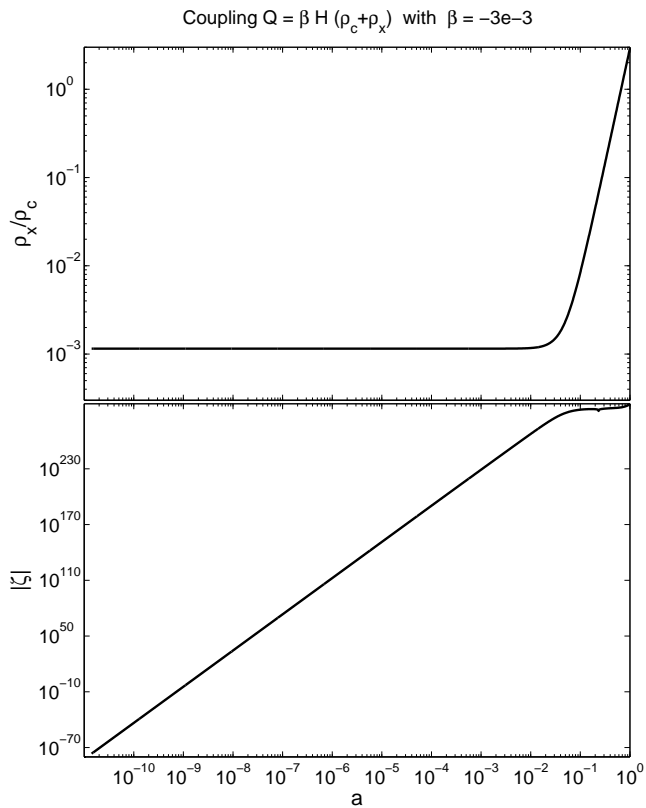}
\caption{The evolution of the background $\rho_x/\rho_c$ (top
panel), and gauge-invariant curvature perturbation $\zeta$ for a
super-Hubble scale $k=7\times10^{-5}\,$Mpc$^{-1}$ (bottom panel)
for the model with coupling given by Eq.~(\ref{QC}). The figure
shows a full numerical solution for $w_x=-0.87$ and
$\beta=-0.003$, starting with the initial value $\zeta = 10^{-80}$
and ending up with oscillations of amplitude
$|\zeta|>10^{+300}$.\label{fig:modCpert}}
\end{figure*}

Using Eqs.~(\ref{dpxC})--(\ref{vpcC}), we can find the early
radiation solution. The solution is simplified if $|\beta|$ is
small enough, and we find that:
 \bea
\theta_x &=& {4\beta \sqrt{\Omega_{r0}}\, \over 3B
\Omega_{c0}(1+w_x)}\,{[n_+^2+(J+1)n_+ -(J+2)] \over (n_++1)}\,
{k^2\over H_0}\,\psi\,,
\label{vpxC3}\\
\delta_x &=&\left[(1+w_x)(n_+-2)+  {2\beta(B+2\beta) \over
B}\right](k\tau)^{-1}\, {\theta_x \over k}\,. \label{dpxC3}
 \eea
The power-law index is given by
\begin{eqnarray}
n_+ & = & \frac{1}{2(1+w_x)} \left\{ -M
 \left( 1+w_x \right) -2N \phantom{\sqrt{\Big[ M \Big]^2 }} \right.\nonumber\\
&& \left.+\sqrt { \Big[ M  \left( 1+w_x \right) +2N
  \Big] ^{2}-4  \left( 1+w_x \right)  \Big[
\left( M+2 \right)  \left( -2-2 w_x+2 N  \right) -3  \left( 1- w_x
\right)  \left( N -3-3
 w_x \right)  \Big] } \right\}\,,
\end{eqnarray}
where $M= -3 w_x+2 \beta^2/B$ and $N= \beta +2 \beta^2/B$. For
example, if $w_x = -0.87$ and $\beta = -0.003$, then $n_+ =
38.95$, as in the previous model.
 For $|\beta|\ll 1$ and $w_x \sim -1$, we have
 \be
n_+ \sim { 6 \over 1+w_x} \gg 1\,.
 \ee
Hence our analytical solution again shows an instability, and this
is confirmed by numerical solution, as illustrated in
Fig.~\ref{fig:modCpert}.

We have also checked numerically that the instability persists
with negligible changes if we make alternative choices of
$Q^\mu_c=-Q^\mu_x$. The two obvious choices are to align
$Q^\mu_c=-Q^\mu_x$ along the dark energy four-velocity or along
the ``centre-of-mass" four-velocity in the dark sector:
 \be
aQ^{\mu}_c =-aQ^\mu_x= -\beta \h (\rho_c + \rho_x) u^{\mu}_{x}\,,
~~~~ aQ^{\mu}_c =-aQ^\mu_x= -\beta \h (\rho_c + \rho_x)
u^{\mu}_{cx}\,,
 \ee
where
\begin{equation}
(\rho_c +\rho_x)u^{\mu}_{cx}= \rho_c u^{\mu}_c + \rho_x u^{\mu}_x
\,.
\end{equation}

\section{Conclusions}

We have given a detailed general analysis of the relativistic
perturbations for a cosmology with coupled dark energy and dark
matter fluids, paying particular attention to the non-adiabatic
features in the dark energy sound speed and to the correct
three-momentum transfer in the dark sector. We specialized to the
case of a constant dark energy equation of state $w_x$, and with
energy-momentum transfer four-vector of the form
$Q^\mu_c=-Q^\mu_x= -\Gamma \rho_c u_c^\mu$, where $\Gamma$ is the
constant interaction rate. We were able to find the fastest
growing early-radiation regular solution of the perturbation
equations to leading order in $k\tau$, and this solution shows a
strong blow-up of the gauge-invariant curvature perturbation on
super-Hubble scales. Even if adiabatic initial conditions are
chosen, the dominant mode quickly overwhelms the adiabatic mode.

The instability occurs no matter how weak the coupling is.
Decreasing $|\Gamma|$ only shifts the blow-up to earlier times.
This is in contrast with the strong-coupling instabilities
discussed in~\cite{Koivisto:2005nr,Bean:2007nx}. The instability
is also non-adiabatic, since growth of $|\zeta|$ is driven by the
total non-adiabatic pressure. This is also in contrast to the
adiabatic instabilities of~\cite{Koivisto:2005nr,Bean:2007nx}.

The origin of this large-scale non-adiabatic instability is not
simply the fact that the dark energy fluid is non-adiabatic, i.e.,
$c_{ax}^2 \neq c_{sx}^2$. In the uncoupled case, the same
non-adiabatic fluid behaviour is also present, but there is no
instability~\cite{Gordon:2004ez}. The coupling plays an essential
role in the large-scale non-adiabatic instability. It appears that
the key driver for the instability is the coupling term that
enters the non-adiabatic dark energy pressure perturbation $\delta
P_x$. This leads to a runaway growth of the dark energy velocity,
in order to maintain momentum balance in the presence of
energy-momentum transfer between the perturbed dark fluids.

We also showed that the instability is not specific to the
coupling model that we introduced. Similar coupling models show
the same qualitative behaviour. We proposed and analyzed covariant
forms of $Q^\mu_c=-Q^\mu_x$ that reduce in the background to two
previously studied coupling models, $aQ=\alpha \h \rho_c$ and
$aQ=\beta \h (\rho_c+\rho_x)$, and showed that in both cases the
non-adiabatic large-scale instability is present. The
perturbations in the second case were previously considered
in~\cite{Olivares:2005tb,Olivares:2006jr}, but they omitted the
coupling term in $\delta P_x$ and this inadvertently removes the
instability. We confirmed numerically that different choices of
$Q^\mu_c=-Q^\mu_x$, i.e., aligning them along different
four-velocities, has a negligible effect on the blow-up of the
perturbations.

Uncoupled models with constant $w_x$ and $c_{sx}^2=1$ are
perfectly well behaved. But it appears that these models are
unstable to the inclusion of coupling, at least for simple forms
of coupling. What are the ways to avoid this instability? We can
relax the assumption $w_x'=0$, using a quintessence model instead
of a fluid model for dark energy. Or more simply, we can use the
parametrization $w_x=w_0+(1-a)w_a$. In this parametrization, $w_x$
is effectively constant, $w_x=w_0+w_a$, in the radiation and early
matter eras. Therefore, our analysis applies unmodified also to
this case.  This may place a tight
theoretical lower bound on $w_a$, since $w_0$ is the late-time
value of $w_x$.
Whether we can produce good fits to CMB and matter power data is
under investigation~\cite{prep}.

The results of this paper, together with previous results on
adiabatic strong-coupling instabilities, strongly constrain the
model space for coupled dark energy. Further constraints on
certain models arise from the background
dynamics~\cite{Amendola:2006qi} and from ``fifth-force"-type
limits in cases where couplings extend to non-dark particles (e.g.
in the case of supergravity-based
quintessence~\cite{Martin:2008qp}).

\[ \]
{\bf Acknowledgements:} We thank Luca Amendola, Bruce Bassett,
Marco Bruni, Sirichai Chongchitnan, Rob Crittenden, Ruth Durrer,
Tomi Koivisto, Claudia Quercellini, Domenico Sapone, Shinji
Tsujikawa and David Wands for useful comments and discussions. JV
and RM are supported by STFC. JV is also supported by the Academy
of Finland.

{}

\end{document}